\documentclass[lettersize,journal]{IEEEtran}

\usepackage{amsmath,amsfonts}
\usepackage{hyperref}
\usepackage{algorithmic}
\usepackage{algorithm}
\usepackage{array}
\usepackage[caption=false,font=normalsize,labelfont=sf,textfont=sf]{subfig}
\usepackage[svgnames]{xcolor}
\usepackage{textcomp}
\usepackage{pgfplots}
\usepackage{pgfplotstable}
\usepackage{stfloats}
\usepackage{url}
\usepackage{verbatim}
\usepackage{color}
\usepackage{xcolor}
\usepackage{graphicx}
\usepackage{cite}
\usepackage{breqn}
\usepackage{multirow}
\newtheorem{theorem}{Theorem}

\newtheorem{definition}{Definition}
\newtheorem{proof}{Proof}

\DeclareMathOperator*{\argmaxA}{arg\,max}
\usepackage{lipsum} 
\setlipsum{%
  par-before = \begingroup\color{gray},
  par-after = \endgroup
}

\hypersetup{
    colorlinks=true,
    linkcolor=black,   % Keeps internal links black (optional)
    citecolor=blue,    % Makes citations blue
    urlcolor=blue      % Makes URLs blue
}
\begin{document}

%\title{Balancing Efficiency and Sustainability \\
%via Market Equilibrium with Externalities}

\title{Balancing Costs and Utilities in Future Networks \\
via Market Equilibrium with Externalities}

% Balancing Efficiency and Sustainability via Market Equilibrium with Externalities

%Balancing Efficiency and Sustainability: An Externality-Aware Resource Allocation Mechanism

%Market Equilibrium with Externalities: A Sustainable Approach to Resource Allocation and Pricing

\author{Mandar Datar$^{1}$ and Mattia Merluzzi$^{1}$~\IEEEmembership{Member,~IEEE}\\
\vspace{-0.6 cm}
        % <-this % stops a space
%\thanks{This paper was produced by the IEEE Publication Technology Group. They are in Piscataway, NJ.}% <-this % stops a space

\thanks{$^{1}$ M.Datar and M.Merluzzi are with CEA, Leti, Univ. Grenoble, F-38000 Grenoble, France.
        {\small  \textcolor{DarkOrchid}{mandar.datar@cea.fr}},
        {\small  \textcolor{DarkOrchid}{mattia.merluzzi@cea.fr}}}

\thanks{This work was supported by the ANR under the France 2030 program, grants "NF-NAI: ANR-22-PEFT-0003" and "NF-JEN: ANR-22-PEFT-0008"}
}



\maketitle

\begin{abstract}
We study the problem of market equilibrium (ME) in future wireless networks, with multiple actors competing and negotiating for a pool of heterogeneous resources (communication and computing) while meeting constraints in terms of global cost. The latter is defined in a general way but is associated with energy and/or carbon emissions. In this direction, service providers competing for network resources do not acquire the latter, but rather the right to consume, given externally defined policies and regulations. We propose to apply the Fisher market model, and prove its convergence towards an equilibrium between utilities, regulatory constraints, and individual budgets. The model is then applied to an exemplary use case of access network, edge computing, and cloud resources, and numerical results assess the theoretical findings of convergence, under different assumptions on the utility function and more or less stringent constraints.
\end{abstract}

\section{Introduction}
Today, the Information and Communication Technology (ICT) sector represents around $2$-$4$\% of global CO$_2$ emissions, mainly due to manufacturing and energy consumption during operations. This includes devices, wireless network segments, core and data centers, from production to operations. However, projections suggest that this global \textit{footprint} share could increase dramatically in the next future \cite{belkhir2018assessing}. This is mainly due to emerging new services that require not only communication resource, but also computing to be pervasive at all network layers, to process the myriad of data that are constantly generated by sensors, applications and vertical sectors such as virtual reality (VR), augmented reality (AR), live broadcasting, automotive, healthcare, and manufacturing. For all these services and beyond, Artificial Intelligence (AI) and the needed computing resources are becoming as essential for 6G as radio spectrum was for previous network generations. As a result, unlike traditional mobile services, where radio resources are the primary resource, modern network services demand heterogeneous resources, including radio access capacity or communication resources, edge storage memory, and computational resources. This inevitably increases the footprint of networks i.e., the \textit{cost} in a broad sense (in terms of deployed hardware resources and energy consumption), but also potentially contributes to new values in terms of economic and societal aspects, spanning from new business models, up to complex systems management towards safe industrial environments for humans. From a footprint perspective, global electricity demand from data centers is expected to triple between 2020 and 2030 \cite{sachs2024ai}. As a result, 6G must take into consideration radio access network (RAN) \cite{lopez2022survey} \cite{viswanathan2022energy} and computing resources in a holistic fashion. To achieve this vision, legacy energy efficiency metrics (e.g., in terms of energy per bit) fall short due to the potential rebound effects \cite{Brockway2021}, and the described prominent role of computing. As an example, the authors of \cite{abrol2016power} indicate that a 5G network, despite its enhanced energy efficiency in J/bit, due to its larger bandwidth and better spatial multiplexing capabilities, could typically consume over 140 more energy than a 4G one, with a similar coverage area. Notably, this unwanted energy consumption arises from network densification, cloud infrastructure, and user equipment (UE), among others. Clearly, only dedicated policies (internal or external) such as budgeting resources (e.g., in terms of energy and/or CO$_2$) remains a valid way to limit the ICT footprint, to best balance it with the benefits. For instance, carbon quotas imply that future cumulative CO$_2$ must meet a given warming limit, thus they are a finite common global resource that must necessarily be shared among countries, whether through prior agreement or as an emergent property of individually determined national efforts \cite{raupach2014sharing}. In this ecosystem and towards this view, a plethora of actors is involved, from policy makers to technology providers, operators, and vertical sectors exploiting the power of communication and computing networks. 
This opens up new challenges in network optimization, including deployment and heterogeneous resource orchestration while balancing performance, efficiency, sustainability, and fair resource distribution. In these contexts, markets are seen as key mechanisms for efficient resource allocation and balancing the interests of all parties, and they have gained significant attention in recent years \cite{buyya2008market,nguyen2019market,datar2023fisher,moro2021joint}.

%This opens up new challenges in network optimization, including deployment and heterogeneous resource orchestration and allocation (communication and compute). 

% As already proposed in the literature \textcolor{red}{[REF]}, we tackle the problem of resource allocation in next generation networks with economic theory tools from market equilibrium. 

%This holistic and dynamic negotiation should target the positive balance between costs and benefits, and it requires continuous interaction among stakeholders and resource consumption budgets. 

%In economic theory, markets serve as the fundamental mechanism that enables and coordinates these exchanges. The basic principles governing these exchanges revolve around the concept of equilibrium and have been extensively studied in the literature for resource allocation, where equilibrium is typically considered as the set of prices at which the supply and demand of all goods in the market are in balance.
%where equilibrium is typically considered as the set of prices at which the supply and demand of all goods in the market are in balance.

In this work, we introduce a novel resource allocation
and pricing scheme that builds on market equilibrium theory. Typically, this is done via a set of prices at which the supply and demand of all goods in the market are in balance. Diverging from these conventional approaches, our mechanism incorporates both values and the external costs associated with resource consumption, such as pollution, energy consumption, and environmental degradation. In particular, we build on the Fisher market model, a special case of the Arrow-Debreu market \cite{arrow1954existence}, where buyers (agents) possess a single commodity used as currency or a fixed budget to acquire different resources (such as bandwidth, CPU, RAM, and storage). Whereas, market operators set prices to ensure resource capacity constraints are met. Eisenberg and Gale \cite{eisenberg1959consensus}, later extended by Jain et al.  \cite{jain2007eisenberg}, showed that when buyers’ utilities are continuous, concave, and homogeneous of degree one (CCH),
%\footnote{A function \( f(x_1, x_2, \dots, x_n) \) is homogeneous of degree \( k \) if for any \( t > 0 \):  
%\(
%f(t x_1, t x_2, \dots, t x_n) = t^k f(x_1, x_2, \dots, x_n).
%\)}, 
market equilibrium can be determined via a convex optimization problem.

We extend this model by incorporating \textit{externalities}, such as the environmental impact of energy consumption, into the market dynamics and demonstrate that equilibrium can still be computed via convex optimization. Our approach aligns with the principles of Pigouvian pricing \cite{baumol1972taxation}, where we introduce taxes alongside prices that internalize external costs within the resource allocation mechanism, ensuring that both capacity constraints and external costs are satisfied, leading to a more sustainable and efficient allocation of resources.

\section{System Model}
We consider a network comprising a set of agents, denoted as \( \mathcal{N} = \{1, 2, \dots, N\} \), each of them requesting a share of divisible resources, represented by \( \mathcal{K} = \{1, 2, \dots, K\} \). Each agent \( n \) specifies its resource allocation preferences through a vector \( \mathbf{x}_n = (x_{n1}, x_{n2}, \dots, x_{nK}) \), where \( x_{nk} \) denotes the quantity of resource \( k \) requested by agent \( n \). The resource set \( \mathcal{K} \) encompasses various types of resources, such as radio spectrum, CPU units, RAM, and storage, which may be distributed across different locations, including edge/fog, and cloud infrastructures, public or private. The overall resource allocation preferences of all agents are collectively represented as \(\mathbf{x} := (\mathbf{x}_n)_{n=1}^{N}.\) In this work, we consider a general setting where agents may represent service providers, network slice tenants, or other entities. 

\subsection{Utility and costs}
Resources generate utilities to the agents, and costs for them but also globally. For example, a given resource has a monetary cost (cost for the agent) and consumes energy and generates carbon emissions (global cost). Throughout the paper, we use the terms \textit{utility} and \textit{cost} to denote positive and negative aspects related to the allocation of resources. Then, we consider CO$_2$ emissions as a global cost representative of today's challenges. Since (during operations) most of these emissions are due to electricity use, we focus on the energy consumption associated with resource utilization to account for the carbon footprint of communication networks.

Each agent $n$ is associated with a utility function denoted as $U_{n}(\mathbf{x}_n): \mathbb{R}^{K} \rightarrow \mathbb{R}_{+}$, indicating the quantified value or utility accrued by the agent $n$ upon acquiring $\mathbf{x}_{n}$ resources.
%We recognize that the utilization of resources in communication networks contributes, directly or indirectly, to CO\(_2\) emissions . Since most of these emissions are due to electricity use, we focus on the energy consumption associated with resource utilization to account for the carbon footprint of communication networks. 
Also, let \(\overline{x}_k\) denote the total utilization of resource \( k \), given by  \(
\overline{x}_k = \sum_{n} x_{nk}.
\) and \(
\overline{\mathbf{x}} =  (\overline{x}_k)_{k=1}^{K}.
\)
We assume that energy consumption depends on overall resource usage, which can be influenced by individual resource consumption or by simultaneous use of multiple resources, for example CPU and RAM together. 

We assume that the total energy consumption considering the carbon intensity of electricity generation is subject to regulatory constraints (or, global budget). These global budget ensure that energy usage and associated CO$_2$ emissions remain within predefined limits set by regulatory authorities. Moreover, these constraints may vary based on location and resource type; for instance, specific regions may impose maximum energy consumption limits to regulate resource usage effectively. This can possibly vary in time and across regions due to the heterogeneous energy mix at country or regional/local scale thanks to, e.g., private grids. The set of regulatory constraints is denoted by $\mathcal{I}$ and defined as 
\begin{align}
    e_{i}(\overline{\mathbf{x}})\leq E_i,\; \forall i \in \mathcal{I}
\end{align}
We assume that certain resources are limited and subject to capacity constraints, while others are virtually unlimited. For instance, resources available at the edge facilities may have restricted capacity, whereas cloud resources can generally be considered to have nearly infinite capacity. Also, cloud resources might be strategically placed in low-carbon energy mix regions, while edge facilities are subject to the location they are deployed at. We denote capacity constraints as follows:
\begin{align}
    \sum\nolimits_n  x_{nk} \leq C_k, k \in \mathcal{K}
\end{align}

 \section{Resource Allocation and Pricing Problem}
%In this work, we focus on the allocation of heterogeneous resources to multiple agents while adhering to energy consumption constraints and resource capacity limitations. 
A key challenge is efficiently distributing resources among competing agents with diverse characteristics and preferences, while ensuring service priority and fairness. To address this, we propose a novel market-based framework designed to balance the interests of all market participants, ensuring that each service is satisfied with its allocation while maintaining high overall utility under different constraints including resources and their cost (e.g., CO$_2$). Toward this goal, we assume that each agent \( n \in \mathcal{N} \) is allocated a budget \( B_n \), which represents artificial currency or \textit{carbon credits} as an example. This budget, determined by a regulatory authority or the agents themselves, may depend on factors such as the agent's service priority as value generator, or its efforts to reduce carbon emissions, for instance, through the adoption of renewable energy sources. The overall goal of externally defining a budget is to properly balance positive and negative impacts of a service. This allocated budget enables agents to procure the resources needed for the services they offer. On the other hand, the market operator sets the prices for the resources to maximize the usage of the resources and drive their distribution towards the target allocation. Given the prices $ \mathbf{p}=(p_1,\dots,p_K) $ set by the market operator, with $p_k$ denoting the price per unit of resource $k$, the feasible demand set of agent $n$ is defined as the set of demands satisfying its budget:
\begin{align}
\mathcal{X}_{n}(\mathbf{p})= \left \{  \mathbf{x}_n \mid  \mathbf{x}_{n}\in \mathbb{R}^{K}, \sum\nolimits_{k}{x_{nk} p_{k}}= B_{n} \right \}\label{feasibledemand}
\end{align} 
We assume that agents exhibit rational self-interested behavior, with each agent seeking to maximize their individual utility. Therefore, the decision problem for each agent $n$ is given as 
\begin{equation}
\begin{aligned}
P_n: \quad & \underset{\mathbf{x}_n\geq 0}{\text{maximize}} & &  U_n(\mathbf{x}_n) \\
& \text{subject to} & & 
\mathbf{x}_n \in \mathcal{X}_{n}(\mathbf{p}), 
\end{aligned}\label{Best_response}
\end{equation}
In this work, our goal is to design a resource allocation mechanism based on a market framework, where we focus on setting the prices in such a way that each agent acts selfishly by requesting resources. The total demand for resources shall then satisfy both the energy constraints and the resource capacity limits. Formally we define market as 
$$\mathcal{M}:=\!\left \langle \mathcal{N},\left(\mathbf{x}_{n}\in\mathbb{R}^K\right)_{n\in\mathcal{N}},\left(U_{n}\right)_{n\in\mathcal{N}},\left(B_{n}\right)_{n\in\mathcal{N}},\mathbf{p}\in\mathbb{R}^K \!\right \rangle, $$ 
The goal of the market is to find an equilibrium between the individual objectives in \eqref{Best_response}, under the regulatory and individual budget constraints. %We assume that 

%\subsection{Market Equilibrium }
\begin{definition}
A market equilibrium (ME) or competitive equilibrium (CE) in the market $\mathcal{M}$ is a pair $(\mathbf{p}^*, \mathbf{x}^*)$ that represents the equilibrium prices and allocation. At this equilibrium, the total energy constraints and resource capacities are met, and agents receive their most preferred resource bundles. Formally, $(\mathbf{p}^*, \mathbf{x}^*)$ constitutes a ME if the following two conditions hold:\\
%\begin{description}
\textbf{C1} Given the resource price vector, every agent $n$ spends its budget so as to receive resource bundle ${\mathbf{x}_{n}}^{*}$ that maximizes its utility:
\begin{align}
{\mathbf{x}_{n}}^{*}\in \argmaxA_{\mathbf{x}_{n}} \left \{  U_{n}(\mathbf{x}_n) \mid  \mathbf{x}_{n} \in \mathcal{X}_n(\mathbf{p}^*) \right \}\;\forall n \in \mathcal{N}\tag{C1.1}\label{bestdemand}
\end{align}\label{C1}
\textbf{C2}  Either energy consumption due to resource usage meets the  restrictions and total resource usage meets capacity  and will be positively priced; otherwise, that resource  has zero price, i.e., we have:
\begin{align}
   \hspace{2cm} &p_{k}^*  = \gamma_{k}^* + \sum_{i\in \mathcal{I}} \lambda_{ik}^*, \quad && \forall k \in \mathcal{K} \tag{C2.1}\label{C2.1}\\
    & \lambda^{*}_{ik} = \nabla_{k} e_{i}(\overline{\mathbf{x}}^*) \lambda^{*}_i, \quad && \forall k \in \mathcal{K}\tag{C2.2} \\ 
    & \lambda^{*}_i \left( e_i(\overline{\mathbf{x}}^*) - E_{i} \right)  = 0, \quad && \forall i \in \mathcal{I} \tag{C2.3}\\
    & \gamma^{*}_k \left( \sum_n  x_{nk}^* - C_k \right)  = 0, \quad && \forall k \in \mathcal{K} \tag{C2.4}\label{C2.4}
\end{align}\label{c2}

%\end{description}
\end{definition}
In the expression $p_{k}^* = \gamma_{k}^* + \sum_{i\in \mathcal{I}} \lambda_{ik}^*$, the term $\gamma_{k}^*$ represents the actual price arising from capacity constraints, while $\lambda_{ik}^*$ can be interpreted as an additional tax (Pigovian price), imposed due to various cost restrictions. Next, we show that the equilibrium solution to the market $\mathcal{M}$ can be computed by solving the convex optimization problem.
\subsection{Equilibrium Solution}
\begin{theorem}\label{thm1}
If each agent's utility function within market $\mathcal{M}$ is continuous, concave, and homogeneous of degree one (CCH), and $e_{i}(\overline{\mathbf{x}})$ is convex and increasing function in $\overline{\mathbf{x}}$, $\forall i \in \mathcal{I}$, the market equlibrium can be computed by solving an optimization program \eqref{EG_Program}
where the optimal allocation $\mathbf{x}^{*}$ and the corresponding prices $\mathbf{p}^*$ (dual variables associated with constraints of the following optimization program as in \eqref{C2.1}-\eqref{C2.4}) represent the ME:
\begin{equation}
    \begin{aligned}
& \underset{\mathbf{x}\geq 0}{\text{maximize}} & &  \sum\nolimits_{n\in\mathcal{N}}B_n\log\left(U_n(\mathbf{x}_n)\right) \\
& \text{subject to} &&\\
&(\lambda_i) && e_{i}(\overline{\mathbf{x}})\leq E_i,\; \forall i \in \mathcal{I}\\
&(\gamma_k)&&\sum\nolimits_{n\in\mathcal{N}}  x_{nk} \leq C_k, k \in \mathcal{K} ,
\end{aligned}\label{EG_Program}
\end{equation}
\end{theorem}
where $\lambda_i, \forall i\in\mathcal{I}$ and $\gamma_k, \forall k\in\mathcal{K}$ denote the Lagrange multipliers associated with the constraints.
\begin{proof}
Consider Lagrangian for optimization program \eqref{EG_Program}
\begin{dmath*}
\mathcal{L}(\mathbf{x}, \boldsymbol{\lambda}, \boldsymbol{\gamma}) = \sum_{n\in\mathcal{N}} B_n \log\left(U_n(\mathbf{x}_n)\right) - \sum\nolimits_{i \in \mathcal{I}} \lambda_i \left( e_{i}(\overline{\mathbf{x}}) - E_i \right) - \sum_{k \in \mathcal{K}} \gamma_k \left( \sum\nolimits_n x_{nk} - C_k \right)+\nu_{nk}x_{nk}.
\end{dmath*}
Let us apply the Karush–Kuhn–Tucker (KKT) conditions \cite{Boyd_Vandenberghe_2004}.
\subsubsection*{Stationarity Conditions}
$\forall k \in \mathcal{K}, \forall n\in\mathcal{N}$, the derivative of the Lagrangian with respect to \( x_{nk} \) is zero at the optimum:  
\[
\left[\frac{B_n \nabla_{nk} U_{n}(\mathbf{x}_n)}{U_n(\mathbf{x}_n)} \right]_{\mathbf{x}=\mathbf{x}^*} - \sum_{i \in \mathcal{I}} \lambda^*_i \left[\nabla_{k} e_{i}(\overline{\mathbf{x}})\right]_{\mathbf{x}=\mathbf{x}^*} - \gamma^*_k +\nu^*_{nk}  = 0.
\]
If we replace $ \sum_{i \in \mathcal{I}} \lambda^*_i \left[\nabla_{k} e_{i}(\overline{\mathbf{x}})\right]_{\mathbf{x}=\mathbf{x}^*} - \gamma^*_k = p^*_k, \; \forall k\in \mathcal{K}$ as defined in \eqref{C2.1} we get 
\begin{equation}
  \left[\frac{B_n \nabla_{nk} U_{n}(\mathbf{x}_n)}{U_n(\mathbf{x}_n)} \right]_{\mathbf{x}=\mathbf{x}^*} -p^*_k+\nu^*_{nk}=0, \quad \forall n\in\mathcal{N}, \forall k \in \mathcal{K}. \label{EG_stationarity}
\end{equation}

\subsubsection*{Primal Feasibility} $\forall n\in\mathcal{N},\forall k \in \mathcal{K},  \forall i \in \mathcal{I},$
\begin{align*}
&x_{nk}^*\geq 0, \quad  e_{i}(\overline{\mathbf{x}}^*) \leq E_i,\quad 
\sum\nolimits_n x^*_{nk} \leq C_k,
\end{align*}

\subsubsection*{Dual Feasibility} $\forall n\in\mathcal{N},\forall k \in \mathcal{K},  \forall i \in \mathcal{I},$
\[
\lambda^*_i\geq 0,  \quad \gamma^*_k \geq 0,  \quad \nu^*_{nk} \geq 0, 
\]

\subsubsection*{Complementary Slackness} $\forall n\in\mathcal{N},\forall k \in \mathcal{K},  \forall i \in \mathcal{I},$
\[
\lambda^*_i \left( e_{i}(\overline{\mathbf{x}}^*) - E_i \right) = 0,\;\gamma^*_k \left( \sum\nolimits_{n\in \mathcal{N}} x_{nk}^* - C_k \right) = 0,\; \nu_{nk}^* x^*_{nk}= 0
\]

Now consider the Lagrangian of the decision problem \eqref{Best_response} of each agent $n\in \mathcal{N}$
\begin{dmath}
\mathcal{L}_n(\mathbf{x}_n, \lambda_n, \boldsymbol{\mu}_n) = U_n(\mathbf{x}_n) - \mu_n \left( \sum\nolimits_k p^*_k x_{nk} - B_n \right) - \sum\nolimits_k \xi_{nk} x_{nk}
\end{dmath}

%where, $\mu_n \geq 0$ is the Lagrange multiplier associated with the constraint $\sum_k p_k x_{nk} \leq B_n$ and $\xi_{nk} \geq 0$ are the Lagrange multipliers associated with the non-negativity constraints $x_{nk} \geq 0$. 
Consider the KKT conditions at the best response
\subsubsection*{Stationarity}
\begin{equation}
\left[\nabla_{nk} U_{n}(\mathbf{x}_n)\right]_{\mathbf{x}_n=\mathbf{x}_n^*} - \mu^*_n p_k +\xi^*_{nk} = 0, \quad \forall k.\label{Stationarity_agent}
\end{equation}

\subsubsection*{ Primal feasibility}
\begin{equation*}
\sum\nolimits_k p_k x^*_{nk} \leq B_n, \quad x^*_{nk} \geq 0, \quad \forall k.
\end{equation*}
\subsubsection*{ Dual feasibility}
\begin{equation*}
\mu^*_n \geq 0, \quad \xi^*_{nk} \geq 0, \quad \forall k.
\end{equation*}
\subsubsection*{ Complementary slackness}
\begin{equation*}
\mu^*_n \left( \sum\nolimits_k p_k x^*_{nk} - B_n \right) = 0, \quad \xi^*_{nk} x^*_{nk} = 0, \quad \forall k.
\end{equation*}
Consider the Stationarity condition \eqref{Stationarity_agent}, multiplying both sides by $x^*_{nk}$ and summing over all $k\in\mathcal{K}$, we have 
\begin{equation}
\left[\nabla_{nk} U_{n}(\mathbf{x}_n)\right]_{\mathbf{x}_n=\mathbf{x}_n^*}x^*_{nk} - \mu_n p_k x^*_{nk}+\xi^*_{nk}x^*_{nk} = 0, \quad \forall k.
\end{equation}
\begin{equation}
\sum\nolimits_k\left[\nabla_{nk} U_{n}(\mathbf{x}_n)\right]_{\mathbf{x}_n=\mathbf{x}_n^*}x^*_{nk} - \mu^*_n B_n = 0, \quad \forall k.
\end{equation}
If the utility function \(U_n\) of each agent \(n \in \mathcal{N}\) is homogeneous of degree 1, then by Euler's homogeneous function theorem, we have:
$\sum\nolimits_k\nabla_{nk} U_{n}(\mathbf{x}_n)x_{nk}= U_n(\mathbf{x}_n)$, replacing \(\mu_n=\frac{U_n(\mathbf{x}_n)}{B_n}\) in \eqref{Stationarity_agent}
\begin{equation}
 \left[\frac{B_n \nabla_{nk} U_{n}(\mathbf{x}_n)}{U_n(\mathbf{x}_n)} \right]_{\mathbf{x}=\mathbf{x}^*} - p_k +\overline{\xi}^*_{nk} = 0, \quad \forall n, k.  \label{statonarity_after_euler}
\end{equation}
where $\overline{\xi}^*_{nk}=\xi^*_{nk}/ \mu^*_n$. 
From \eqref{EG_stationarity} and \eqref{statonarity_after_euler}, when $\mathbf{p}=\mathbf{p}^*$ the KKT conditions at the optimum of problem \eqref{EG_Program} and the set of KKT conditions at the best response of the agents' decision problem \eqref{Best_response} are equal, thus establishing the claim.
\end{proof}
\subsection{Nash welfare and fairness}
We proved that the market equilibrium for $\mathcal{M}$ can be determined by solving the optimization program in \eqref{EG_Program}. Building on the work of \cite{eisenberg1959consensus}, we assert that market equilibrium allocations maximize Nash welfare. This follows from the fact that optimizing the objective function in \eqref{EG_Program} is equivalent to maximizing the Nash welfare function.
\begin{equation}
\argmaxA_{\mathbf{x}\in\mathcal{X}}\Pi_n U_{n}(\mathbf{x}_n)^{B_n}=\argmaxA_{\mathbf{x}\in\mathcal{X}}\sum\nolimits_{n\in\mathcal{N}}B_n\log\left(U_n(\mathbf{x}_n)\right)
\end{equation}

Further, considering the first order condition for convex optimization 
\begin{equation}
     \sum_n B_n\frac{U_n(\mathbf{x}_n)-U_n(\mathbf{x}^*_n)}{U_n(\mathbf{x}^*_n)}\leq 0\quad \forall \;\mathbf{x},\mathbf{x}^*\in \mathcal{X}
\end{equation}
we observe that the equilibrium utility vector \(\left(U_1(\mathbf{x}^*_1),\dots, U_N(\mathbf{x}^*_N)\right)\)
can be interpreted as proportionally fair. Specifically, when comparing this equilibrium vector to any other feasible utility vector  \(\left(U_1(\mathbf{x}_1),\dots, U_N(\mathbf{x}_N)\right)\), the sum of the proportional changes in utilities is non-positive. This implies that any deviation from the equilibrium utility allocation leads to a net negative or zero aggregate proportional change across all users. Therefore, the proposed allocation scheme maximizes Nash welfare or achieves the proportional fairness criteria while distributing the resources among the agents.

\subsection{Examples of utility}
One of the examples of a general class of CCH utility functions is the CES (constant elasticity of substitution) utility function,  which is represented as
\(
u_n(\mathbf{x}_n) = \left(\sum_k v_{nk}(x_{nk})^{\rho}\right)^{1/\rho},
\)
where \(\rho\) defines the function's elasticity. The parameter \(\rho\) can range from \((-\infty, 0) \cup (0, 1]\). When \(\rho = 1\), the utility is linear, representing perfect substitutes. As \(\rho \to 0\), the utility function becomes Cobb-Douglas, which has a multiplicative form $u_{n}(\mathbf{{x}_n})=\Pi_{k} (x_{nk})^{v_{nk}}$. As \(\rho \to -\infty\), it becomes Leontief  $u_{n}(\mathbf{{x}_n})=\underset{k}{\text{min}} \lbrace\frac{x_{nk} }{v_{nk}}\rbrace$, representing perfect complements. Linear valuation signifies that resources are perfect substitutes, whereas Leontief indicates perfect complements. The CES function interpolates between perfect substitutes and complements on the basis of the value \(\rho\). Moreover, applying the CES form to an existing CES utility function, such as \(
\left(\sum_k v_{n}(u_n(\mathbf{x}_n))^{\rho}\right)^{1/\rho},
\) also results in a CCH function, allowing for a broad range of utility representations. For Leontief utilities, the derivative is $ \nabla_{nk}U_n(\mathbf{x}_n)=\frac{1}{v_{nk}}$ if $k$ is binding constraint otherwise its value its $0$, multiplying by \( x_{nk} \) and summing, $\sum_{k=1}^{K} x_{nk} \nabla_{nk}U_{n}(\mathbf{x}_n) = x_{nk} \frac{1}{v_{nk}} = U_n(\mathbf{x}_n).$ Thus, Euler’s theorem holds for the Leontief utility function.

\section{Numerical Experiments}
For our numerical experiment and to better interpret outcomes, we consider a scenario with two service providers (SPs) competing for resources in a specific location. The first provider supports a radio-intensive application, while the second supports a computing-intensive application. The area has wireless communication resources and an edge computing facility, and is connected to the cloud for additional computing power. Let \({U}^1_n\) and \(U^2_n\) represent the utility of SP \(n\) when using cloud and MEC facility resources, respectively. The total utility is given by \(U_n = {U}_{n,e} + {U}_{n,c}\), with $e$ and  $c$ denoting the edge and cloud, respectively. We assume that both follow Leontief utility functions and represented as \({U}_{n,  e(c)}=\underset{}{\text{min}}\left\{ \frac{x_{nk_1, e(c)}}{d_{nk_1, e(c)}},\frac{x_{nk_2, e(c)}}{d_{nk_2, e(c)}},\frac{x_{nk_3, e(c)}}{d_{nk_3, e(c)}}\right\}\), with $k_1$, $k_2$, $k_3$ being RAM, CPU and bandwidth (BW) resources. Table \ref{basedemandtable} describes the base demand vector for each SP. 
\begin{table}[h]
\centering
\caption{The base demand vector of service classes}
\begin{tabular}{|c|c|c|c|c|c|c|}
\hline 
Service Class & CPU & RAM & BW  \\ 
\hline 
  \multirow{2}{*}{\begin{tabular}{c|c}

\multirow{2}{*}{Radio-Intensive} & MEC \\
                              & Cloud\\
\end{tabular}} & 2 CPUs & 8 GB & 10MHz\\ & 4 CPUs & 8 GB & 15 MHz \\ 
\hline 
 \multirow{2}{*}{\begin{tabular}{c|c}

\multirow{2}{*}{CPU-Intensive } & MEC \\
                              & Cloud \\

\end{tabular}} & 8 CPU & 14 GB & 3 MHz \\ & 10 CPU & 16 GB & 6 MHz\\ \hline 
\end{tabular} 
\label{basedemandtable}
\end{table}

We assume that the primary sources of energy consumption are radio resource usage and CPU resource usage, while the energy cost associated with RAM is negligible. The energy function for each resource Radio and CPU units follows a power form, \(e_k(\overline{x}_k) = (\overline{x}_k)^{\beta_{k}}\), with $\beta_{k}\geq 1$. This allows us to explore the different trade-offs based on the specific energy consumption model. We impose restrictions on energy consumption at two levels:  
(i) total energy consumption across all resources, and  
(ii) local energy consumption that applies specifically to the radio resources and the MEC facility.

\par We employ the proposed resource allocation scheme by solving \eqref{EG_Program} and then compare the ME-based allocation with the social optimum (SO) allocation, which aims to maximize the total weighted sum of services provided by SPs while meeting energy consumption and resource capacity constraints.

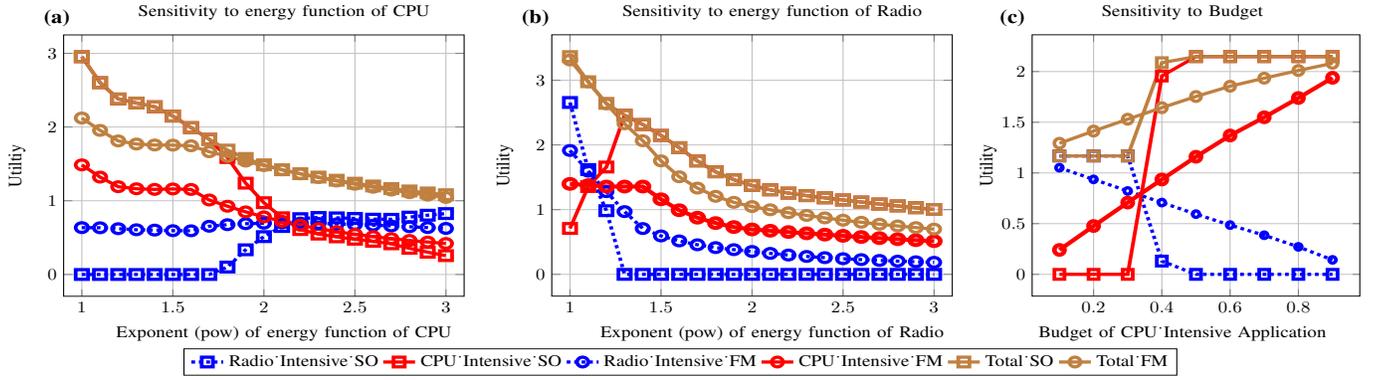
\begin{figure*}
    %\centering
    \resizebox{\textwidth}{5cm}{
   \begin{tikzpicture}
   \node[anchor=south west] at (-2,5.75){\large\textbf{(a)}};
   \node[anchor=south west] at (7,5.75){\large\textbf{(b)}};
    \node[anchor=south west] at (16,5.75){\large\textbf{(c)}};
        \begin{axis}[
             name=ax1,
            title={Sensitivity to energy function of CPU },
            xshift=-1.5cm,
            xmin=0.9, xmax=3.1,
            xscale=1.1, yscale=1,
            xlabel={Exponent (pow) of energy function of CPU},
            ylabel={Utiltiy},
            ylabel style={yshift=-10pt},
            grid=major,
            legend style={at={(1.4,-0.2)}, anchor=north,legend columns=-1},
        ]
           %\addplot table[x=PF_CPU,y=Us0]{Data_files/filtered_1.csv};

           \addlegendentry{Radio_Intensive_SO}
             \addplot[dotted, line width=2pt,mark=square,mark options={solid},mark size=3,blue] table [x index=0, y index=1, col sep=comma] {Data_files/filtered_1.csv};
               \addlegendentry{CPU_Intensive_SO}
             \addplot[line width=2pt,mark=square,mark size=3, red] table [x index=0, y index=2, col sep=comma] {Data_files/filtered_1.csv};
             \addlegendentry{Radio_Intensive_FM}
             \addplot[dotted,line width=2pt,mark=o,mark options={solid},mark size=3, blue] table [x index=0, y index=1, col sep=comma] {Data_files/filtered_2.csv};
            \addlegendentry{CPU_Intensive_FM} 
             \addplot[line width=2pt,mark=o,mark size=3,red] table [x index=0, y index=2, col sep=comma] {Data_files/filtered_2.csv};
\addlegendentry{Total_SO}
\addplot[line width=2pt,mark=square,mark size=3, brown] table [x index=0,  y expr=\thisrow{Us0} + \thisrow{Us1}, col sep=comma] {Data_files/filtered_1.csv};
\addlegendentry{Total_FM}
\addplot[line width=2pt,mark=o,mark size=3, brown] table [x index=0,  y expr=\thisrow{Us0} + \thisrow{Us1}, col sep=comma] {Data_files/filtered_2.csv};
\end{axis}

\begin{axis}[
        name=ax2,
        at={(ax1.south east)},
         xmin=0.9, xmax=3.1,
       xscale=1.1, yscale=1,
        xshift=1.5cm,
            title={Sensitivity to energy function of Radio },
            xlabel={Exponent (pow) of energy function of Radio},
            ylabel={Utility},
            ylabel style={yshift=-10pt},
            grid=major,
        ]
           %\addplot table[x=PF_CPU,y=Us0]{Data_files/filtered_1.csv};
             \addplot[dotted,line width=2pt,mark=square,mark options={solid},mark size=3,blue] table [x index=0, y index=1, col sep=comma] {Data_files/filtered_3.csv};
             \addplot[line width=2pt,mark=square,mark size=3 ,red] table [x index=0, y index=2, col sep=comma] {Data_files/filtered_3.csv};
             \addplot[dotted,line width=2pt,mark=o,mark options={solid},mark size=3, blue] table [x index=0, y index=1, col sep=comma] {Data_files/filtered_4.csv};
             \addplot[line width=2.5pt,mark=o,mark size=3, red] table [x index=0, y index=2, col sep=comma] {Data_files/filtered_4.csv};

    \addplot[line width=2pt,mark=square,mark size=3, brown] table [x index=0,  y expr=\thisrow{Us0} + \thisrow{Us1}, col sep=comma] {Data_files/filtered_3.csv};

\addplot[line width=2pt,mark=o,mark size=3, brown] table [x index=0,  y expr=\thisrow{Us0} + \thisrow{Us1}, col sep=comma] {Data_files/filtered_4.csv};

        \end{axis}

        \begin{axis}[
         at={(ax2.south east)}, xshift=1.5cm,
            title={Sensitivity to Budget},
            xscale=0.9, yscale=1,
            xlabel={Budget of CPU_Intensive Application},
            ylabel={Utility},
            ylabel style={yshift=-10pt},
            grid=major,
        ]
           %\addplot table[x=PF_CPU,y=Us0]{Data_files/filtered_1.csv};
             \addplot [dotted,line width=2pt,mark=square,mark options={solid},mark size=3,blue] table [x index=0, y index=1, col sep=comma] {Data_files/filtered_5.csv};
             \addplot [line width=2pt,mark=square, mark size=3,red] table [x index=0, y index=2, col sep=comma] {Data_files/filtered_5.csv};
             \addplot[dotted,line width=2pt,mark=o,
    mark options={solid}, blue] table [x index=0, y index=1, col sep=comma] {Data_files/filtered_6.csv};
             \addplot[line width=2.5pt,mark=o,mark size=3, red] table [x index=0, y index=2, col sep=comma] {Data_files/filtered_6.csv};

     \addplot[line width=2pt,mark=square,mark size=3, brown] table [x index=0,  y expr=\thisrow{Us0} + \thisrow{Us1}, col sep=comma] {Data_files/filtered_5.csv};

\addplot[line width=2pt,mark=o,mark size=3, brown] table [x index=0,  y expr=\thisrow{Us0} + \thisrow{Us1}, col sep=comma] {Data_files/filtered_6.csv};

        \end{axis}
\end{tikzpicture}}

\caption{Sensitivity of the agents' utility to various factors: (a) the CPU energy function, (b) the Radio energy function, and (c) agents' budgets}
    \label{fig:sensitivity}

\end{figure*}

We analyze the sensitivity of various parameters, such as cost functions and budgets, on SPs' utilities under both allocation schemes. Figs. \ref{fig:sensitivity} (a) and (b) illustrate how the utilities of SPs vary with the degree of the energy cost function, and they are obtained by adjusting the power of the cost function $\beta_k$ from 1 to 3. It can be observed how, as the exponent increases, the utilities of SPs decrease, to meet the constraint. However, SO allocation exhibits high sensitivity to cost variations and results in unfair outcomes, as it disproportionately benefits only one SP. In contrast, the proposed ME-based allocation demonstrates greater stability and fairness in resource distribution. In the Fig. \ref{fig:sensitivity}(c), we analyze how resource allocation is affected by agents' budgets. We vary the budget of the Radio-Intensive SP from 0.1 to 0.9 and plot the utilities of both SPs as the budget varies under both schemes. We observe that under the SO scheme, the agent with a low budget receives very little utility. However, under the proposed ME-based approach, the system achieves better fairness and shows less sensitivity to budget changes thus achieving better balance between cost, utility and faireness.

\par
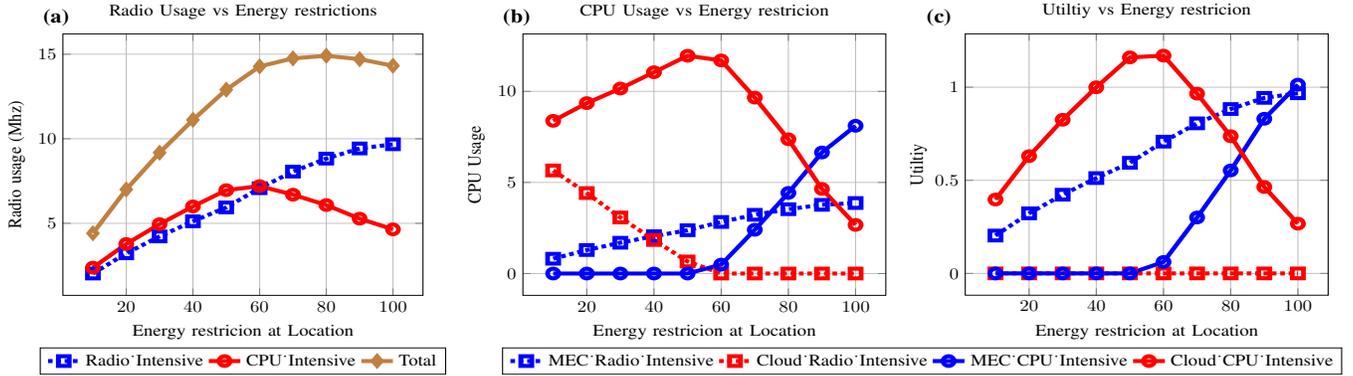
\begin{figure*}
    %\centering
\resizebox{0.33\textwidth}{5cm}{
   \begin{tikzpicture}
    \node[anchor=south west] at (-2,5.75){\large\textbf{(a)}};
        \begin{axis}[
             name=ax1,
            title={Radio Usage vs Energy restrictions},
            xshift=-1.5cm,
            xscale=1, yscale=1,
            xlabel={Energy restricion at Location},
            ylabel={Radio usage (Mhz)},
            ylabel style={yshift=-10pt},
            grid=major,
            legend style={at={(0.5,-0.2)}, anchor=north,legend columns=-1},
        ]
           %\addplot table[x=PF_CPU,y=Us0]{Data_files/filtered_1.csv};

           \addlegendentry{Radio_Intensive}
             \addplot[dotted,line width=2.5pt,mark=square,mark options={solid},mark size=3,blue] table [x index=0, y index=2, col sep=comma] {Data_files/filtered_13.csv};
               \addlegendentry{CPU_Intensive}
             \addplot[line width=2.5pt,mark=o,mark size=3, red] table [x index=0, y index=3, col sep=comma] {Data_files/filtered_13.csv};
             \addlegendentry{Total}
             \addplot[mark=diamond,mark size=3,brown,line width=2.5pt,] table [x index=0, y index=1, col sep=comma] {Data_files/filtered_13.csv};
 \end{axis}

       \end{tikzpicture}}
 \resizebox{0.65\textwidth}{5cm}{
   \begin{tikzpicture}
   \node[anchor=south west] at (-2,5.75){\large\textbf{(b)}};
    \node[anchor=south west] at (6,5.75){\large\textbf{(c)}};
        \begin{axis}[
             name=ax1,
            title={CPU Usage vs Energy restricion},
            xshift=-1.5cm,
            xscale=1, yscale=1,
            xlabel={Energy restricion at Location},
            ylabel={CPU Usage},
            ylabel style={yshift=-10pt},
            grid=major,
            legend style={at={(1.1,-0.2)}, anchor=north,legend columns=-1},
        ]
           %\addplot table[x=PF_CPU,y=Us0]{Data_files/filtered_1.csv};

           \addlegendentry{MEC_Radio_Intensive}
             \addplot [dotted, line width=2.5pt,mark size=3,mark=square,mark options={solid},blue] table [x index=0, y index=2, col sep=comma] {Data_files/filtered_12.csv};
               \addlegendentry{Cloud_Radio_Intensive}
             \addplot[dotted, line width=2.5pt,mark size=3,mark=square, mark options={solid}, red]  table [x index=0, y index=3, col sep=comma] {Data_files/filtered_12.csv};
             \addlegendentry{MEC_CPU_Intensive}
              \addplot[line width=2.5pt,mark size=3,mark=o,
    mark options={solid}, blue] table [x index=0, y index=4, col sep=comma] {Data_files/filtered_12.csv};
            \addlegendentry{Cloud_CPU_Intensive} 
              \addplot[line width=2.5pt,mark size=3,mark=o,
   mark options={solid}, red] table [x index=0, y index=5, col sep=comma] {Data_files/filtered_12.csv};

        \end{axis}

        \begin{axis}[
        name=ax2,
        at={(ax1.south east)},
       xscale=1, yscale=1,
        xshift=1.5cm,
            title={Utiltiy vs Energy restricion  },
            xlabel={Energy restricion at Location},
            ylabel={Utiltiy},
            ylabel style={yshift=-10pt},
            grid=major,
        ]
           %\addplot table[x=PF_CPU,y=Us0]{Data_files/filtered_1.csv};
             \addplot  [dotted, line width=2.5pt,mark=square,mark options={solid},mark size=3,blue] table [x index=0, y index=1, col sep=comma] {Data_files/filtered_10.csv};
            \addplot [dotted, line width=2.5pt,mark=square,mark options={solid},mark size=3, red] table [x index=0, y index=2, col sep=comma] {Data_files/filtered_10.csv};
           \addplot[line width=2.5pt,mark=o,mark size=3, blue] table [x index=0, y index=3, col sep=comma] {Data_files/filtered_10.csv};
             \addplot[line width=2.5pt,mark=o,
    mark size=3, red] table [x index=0, y index=4, col sep=comma] {Data_files/filtered_10.csv};

        \end{axis}
\end{tikzpicture}}

\caption{Variation in the usage of: (a) radio, (b) CPU, and the utility derived from both MEC and cloud facilities vs. local energy consumption constraints. }
    \label{fig:restriction}

\end{figure*}

 Fig. \ref{fig:restriction} illustrates how restrictions to local energy consumption impact resource allocation and utilities. In plots in Fig. \ref{fig:restriction} (a),  we vary the local energy constraints from 10 to 90 units and analyze the usage of radio resources for uploading, as well as MEC and cloud facilities. As restrictions are relaxed, total radio resource usage increases. However, the radio resource usage for cloud uploading initially increases and then decreases. This occurs because loosening the restrictions allows for greater use of MEC resources. As a result, instead of uploading to the cloud, SPs prefer exploiting MEC resources. This observation is further supported by the plots in the Fig.\ref{fig:restriction} (b)-(c), which show that CPU usage at the cloud facility increases when restrictions are eased, as more radio resources become available for cloud uploads. However, as the restrictions are relaxed further, MEC usage increases, reducing the dependency on cloud facilities.   

\section{Conclusions and future directions}
We proposed a new resource allocation and pricing mechanism based on an extended Fisher market model that accounts for external costs such as energy consumption and environmental impact. By incorporating Pigouvian taxes, our approach satisfies resource constraints towards sustainability while maximizing Nash welfare and proportional fairness among agents. We showed that the equilibrium problem can be solved using convex optimization, demonstrating the low complexity, practicality and effectiveness of our method.
%\printbibliography
\bibliography{sample-bibliography}
\bibliographystyle{IEEEtran}

\end{document}